\documentclass[floatfix,aps,twocolumn,prl,superscriptaddress, ]{revtex4}
\usepackage{amsmath}
\usepackage{amssymb}
\usepackage{graphicx}
\usepackage{dcolumn}
\usepackage{natbib}
\usepackage{bm}
\usepackage{epsfig}

\begin{document}

\title{Band structure asymmetry of bilayer graphene revealed by infrared
spectroscopy}
\author{Z. Q. Li}
\email{zhiqiang@physics.ucsd.edu}
\affiliation{Department of Physics, University of California, San Diego, La Jolla,
California 92093, USA}
\author{E. A. Henriksen}
\affiliation{Department of Physics, Columbia University, New York, New York 10027, USA}
\author{Z. Jiang}
\affiliation{Department of Physics, Columbia University, New York, New York 10027, USA}
\affiliation{National High Magnetic Field Laboratory, Tallahassee, Florida 32310, USA}
\author{Z. Hao}
\affiliation{Advanced Light Source Division, Lawrence Berkeley National Laboratory,
Berkeley, California~94720, USA}
\author{M. C. Martin}
\affiliation{Advanced Light Source Division, Lawrence Berkeley National Laboratory,
Berkeley, California~94720, USA}
\author{P. Kim}
\affiliation{Department of Physics, Columbia University, New York, New York 10027, USA}
\author{H. L. Stormer}
\affiliation{Department of Physics, Columbia University, New York, New York 10027, USA}
\affiliation{Department of Applied Physics and Applied Mathematics, Columbia University,
New York, New York 10027, USA}
\affiliation{Bell Labs, Alcatel-Lucent, Murray Hill, New Jersey 07974, USA}
\author{D. N. Basov}
\affiliation{Department of Physics, University of California, San Diego, La Jolla,
California 92093, USA}
\date{\today }

\begin{abstract}
We report on infrared spectroscopy of bilayer graphene integrated in gated
structures. We observed a significant asymmetry in the optical conductivity
upon electrostatic doping of electrons and holes. We show that this finding
arises from a marked asymmetry between the valence and conduction bands,
which is mainly due to the inequivalence of the two sublattices within the
graphene layer. From the conductivity data, the energy difference of the two
sublattices is determined.
\end{abstract}

\maketitle

\bigskip Recently there has been unprecedented interest in carbon-based
materials due to the discovery of graphene \cite{PheneNature}. Among all
carbon systems, bilayer graphene stands out due to its remarkable properties
such as the formation of a tunable band gap between the valence and
conduction bands \cite{Castro}\cite{Ohta}\cite{Gap}\cite{McCann}: a property
not attainable in common semiconductors. The vast majority of previous
experimental and theoretical studies of bilayer graphene assumed a symmetric
band structure that is governed by the interlayer coupling energy $\gamma
_{1}$. This is in contrast with a significant electron-hole asymmetry
observed in cyclotron resonance \cite{Erik} and cyclotron mass experiments 
\cite{Castro}. Several theoretical proposals have been put forward to
explain these results \cite{Castro}\cite{Kusminskiy}. Yet, the microscopic
origin of the observed effects remains unknown.

Here we present the first investigation of the optical conductivity of
bilayer graphene via infrared spectroscopy. We observed dramatic differences
in the evolution of the conductivity for electron and hole polarities of the
gate voltage. We show that small band parameters other than $%
\gamma _{1}$ give rise to an asymmetry between the valence and conduction
bands, in contrast to the commonly assumed symmetric band structure. The
systematic character of our IR data enables us to extract an energy
difference between the A and B sublattices within the same graphene layer
(Fig 1(b)) of $\delta _{AB}\approx $18meV. We analyze some of the
implications of these findings for other properties of bilayer graphene.

Infrared (IR) reflectance R($\omega $) and transmission T($\omega $)
measurements were performed on bilayer graphene samples on SiO$_{2}$/Si
substrate \cite{Erik} as a function of gate voltage V$_{g}$ at 45K employing
synchrotron radiation, as described in \cite{Li08}. We find that both R($%
\omega $) \cite{Wang, Kuzmenko-bilayer} and T($\omega $) spectra of the
bilayer graphene device can be strongly modified by a gate voltage. Figure 1
shows the transmission ratio data at several voltages normalized by data at
the charge neutrality voltage V$_{CN}$: T(V)/T(V$_{CN}$), where V$_{CN}$ is
the voltage corresponding to the minimum DC conductivity, and V= V$_{g}-$V$%
_{CN}$. The T(V)/T(V$_{CN}$) spectra are dominated by a dip at around 3000 cm%
$^{-1}$, the magnitude of which increases systematically with voltage. Apart
from the main dip, a peak was observed in the T(V)/T(V$_{CN}$) data below
2500 cm$^{-1}$, which shifts systematically with voltage. This latter
feature is similar to the T(V)/T(V$_{CN}$) data for single layer graphene 
\cite{Li08}. The gate-induced enhancement in transmission (T(V)/T(V$_{CN}$)$>
$1) below 2500 cm$^{-1}$ and above 3500 cm$^{-1}$ implies a decrease of the
absorption with voltage in these frequency ranges.

The most informative quantity for exploring the quasiparticle dynamics in
bilayer graphene is the two dimensional (2D) optical conductivity $\sigma
_{1}\left( \omega \right) +i\sigma _{2}\left( \omega \right) $ \cite{Li08}%
\cite{LiPRL}. First, we extracted the optical conductivity at V$_{CN}$ from
the reflectance data (not shown) employing a multilayer analysis of the
device \cite{Li08}\cite{LiPRL}. We find that $\sigma _{1}\left( \omega
,V_{CN}\right) $ has a value of $2\ast (\pi e^{2}/2h)$ at high energies,
with a pronounced peak at 3250 cm$^{-1}$ (inset of Fig 2(b)). This
observation is in agreement with theoretical analysis on undoped bilayer
graphene\cite{BilayerIR}\cite{Nilsson}\cite{Nicol}. Our high energy data
agree with recent experiments in the visible region \cite{GeimIR}. The peak
around 3250 cm$^{-1}$ can be assigned to the interband transition in undoped
bilayer near the interlayer coupling energy $\gamma _{1}$.

An applied gate voltage shifts the Fermi energy E$_{F}$ to finite values
leading to significant modifications of the optical conductivity. The $%
\sigma _{1}\left( \omega ,V\right) $ and $\sigma _{2}\left( \omega ,V\right) 
$ spectra extracted from voltage-dependent reflectance and transmission data 
\cite{Li08} are shown in Fig 2. At frequencies below 2500 cm$^{-1}$, we
observe a suppression of $\sigma _{1}\left( \omega ,V\right) $ below $2\ast
(\pi e^{2}/2h)$ and a well-defined threshold structure, the energy of which
systematically increases with voltage. Significant conductivity was observed
at frequencies below the threshold feature. These observations are similar
to the data in single layer graphene \cite{Li08}. The threshold feature
below 2500 cm$^{-1}$ can be attributed to the onset of interband transitions
at 2E$_{F}$, as shown by the arrow labeled e$_{1}$ in the inset of Fig 2(a)
and (b). The observed residual conductivity below 2E$_{F}$ is in contrast to the
theoretical absorption for ideal bilayer graphene \cite{Nilsson}\cite{Nicol}
that shows nearly zero conductivity up to 2E$_{F}$. Similar to single layer
graphene, the residual conductivity may originate from disorder effects \cite%
{Nilsson} or many body intertactions \cite{Li08}. Apart from the above
similarities, the optical conductivity of bilayer graphene is significantly
different from the single layer conductivity. First, the energy range where the
conductivity $\sigma _{1}\left( \omega ,V\right) $ is impacted by the gate
voltage extends well beyond the 2E$_{F}$ threshold. Furthermore, we find a
pronounced peak near 3000 cm$^{-1}$, the oscillator strength of which shows
a strong voltage dependence. This peak originates from the interband
transition between the two conduction bands or two valence bands (inset of
Fig.2a) \cite{Nilsson}\cite{Nicol}.

The voltage dependence of the Fermi energy in bilayer graphene can be
extracted from $\sigma _{2}\left( \omega ,V\right) $ using a similar
procedure as in \cite{Li08}. In order to isolate the 2E$_{F}$ feature, we
fit the main resonance near 3000 cm$^{-1}$ with Lorentzian oscillators
and then subtracted them from the experimental $\sigma _{2}\left( \omega
,V\right) $ spectra to obtain $\sigma _{2}^{diff}\left( \omega ,V\right) $.
The latter spectra reveal a sharp minimum at $\omega $=2E$_{F}$ (Fig 2(c))
in agreement with single layer graphene \cite{Li08}. Figure 3a depicts the
experimental 2E$_{F} $ values along with the theoretical result in \cite%
{McCann}. Assuming the Fermi velocity v$_{F}$ in bilayer graphene is similar
to that in single layer graphene (v$_{F}$=1.1$\times $10$^{6}$ m/s), we find
that our data can be fitted with $\gamma _{1}$=450$\pm $80meV. Equally
successful fits can be obtained assuming the Fermi velocity and interlayer
coupling in the following parameter space: v$_{F}$=1.0-1.1$\times $10$^{6}$
m/s and $\gamma _{1}$=360-450 meV. Previous theoretical and experimental
studies showed that an applied gate voltage opens a gap $\Delta $ between
the valence and conduction bands \cite{Castro}\cite{Ohta}\cite{Gap}\cite%
{McCann}. Because $\Delta (V)$ is much smaller than 2E$_{F}$(V) for any
applied bias in bottom-gate devices \cite{McCann}, it has negligible effects
on the experimentally observed 2E$_{F}$(V) behavior.

The central result of our study is an observation of a pronounced asymmetry
in evolution of the optical conductivity upon injection of electrons or
holes in bilayer graphene. Specifically, the frequencies of the main peak $%
\omega _{peak}$ in $\sigma _{1}\left( \omega ,V\right) $ are very distinct
for E$_{F}$ on the electron and hole sides, as shown in Fig 3(b). In
addition, $\omega _{peak}$ on the electron side shows a much stronger
voltage dependence compared to that on the hole side. All these features are
evident in the raw data in Fig.1, where the resonance leads to a dip in
T(V)/T(V$_{CN}$) spectra. These behaviors are reproducible in multiple gated
samples. Such an electron-hole asymmetry is beyond a simple band structure
only taking $\gamma _{1}$ into account, which predicts symmetric properties
between electron and hole sides.

We propose that the electron-hole asymmetry in our $\sigma _{1}\left( \omega
,V\right) $ data reflects an asymmetry between valence and conduction bands.
Such an asymmetric band structure arises from finite band parameters $\delta
_{AB}$ and v$_{4}$, where $\delta _{AB}$ (denoted as $\Delta $ in \cite%
{Graphite}\cite{Nilsson}) is the energy difference between A and B
sublattices within the same graphene layer, and v$_{4}$=$\gamma _{4}/\gamma
_{0}$. $\gamma _{4}$ and $\gamma _{0}$ are defined as interlayer
next-nearest-neighbor coupling energy and in-plane nearest-neighbor coupling
energy, respectively \cite{Graphite}\cite{Nilsson}. We first illustrate the
effects of $\delta _{AB}$ and v$_{4}$ on the energy bands of bilayer
graphene E$_{i}$(k) (i=1,2,3,4), which can be obtained from solving the
tight binding Hamiltonian Eq (6) in Ref. \cite{Nilsson}. We find that finite
values of\ $\delta _{AB}$ and v$_{4}\ $break the symmetry between valence
and conduction bands, as schematically shown in the inset of Fig 2(a).
Specifically, $\delta _{AB}$ induces an asymmetry in E$_{1}$ and E$_{4}$
bands such that E$_{1}>$-E$_{4}$ at k=0, whereas v$_{4}$ induces an
electron-hole asymmetry in the slope of the valence and conduction bands.
With finite v$_{4}$, the bands E$_{1}$ and E$_{2}$ are closer and E$_{3}$
and E$_{4}$ are further apart at high k compared to those with zero v$_{4}$
value.

Next we examine the effects of $\delta _{AB}$ and v$_{4}$ on $\sigma
_{1}\left( \omega ,V\right) $. It was predicted theoretically \cite{Nicol}
that the main peak in $\sigma _{1}\left( \omega ,V\right) $ occurs in the
frequency range between two transitions labeled e$_{2}$ and e$_{3}$ as shown
in the inset of Fig 2(a) and (b). Here e$_{2}$=$-$E$_{4}$(k=0)$-\Delta /2$
and e$_{3}$=E$_{3}$(k=k$_{F}$)$-$E$_{4}$(k=k$_{F}$) for the hole side, and e$%
_{2}$=E$_{1}$(k=0)$-\Delta /2$ and e$_{3}$=E$_{1}$(k=k$_{F}$)$-$E$_{2}$(k=k$%
_{F}$) for the electron side\cite{Nicol}, with $\Delta $ defined as the gap
at k=0. For zero values of $\delta _{AB}$ and v$_{4}$, e$_{2}$ and e$_{3}$
transitions are identical on the electron and hole sides. The finite values
of $\delta _{AB}$ and v$_{4}$ induce a significant inequality between e$_{2}$
and e$_{3}$ on the electron and hole sides. We first focus on the low
voltage regime, where $\omega _{peak}$=e$_{2}$=e$_{3}$. Because v$_{F}$ and v%
$_{4}$ always enter the Hamiltonian in the form of v$_{F}$k and v$_{4}$k
products\cite{Nilsson}, these terms give vanishing contributions at low V,
where k goes to zero. Consequently, $\omega _{peak}$ value at low bias is
solely determined by $\gamma _{1}$ and $\delta _{AB}$, with $\omega _{peak}$=%
$\gamma _{1}+\delta _{AB}$ and $\omega _{peak}$=$\gamma _{1}-\delta _{AB}$
for the electron and hole sides, respectively. At V$_{CN}$ (0V), interband
transitions between the two conduction bands and the two valence bands are
both allowed, which leads to a broad peak centered between $\gamma
_{1}+\delta _{AB}$ and $\gamma _{1}-\delta _{AB}$ (Fig 3(b)). From the two
distinct low voltage $\omega _{peak}$ values on the electron and hole sides
shown in Fig 3(b), the values of $\gamma _{1}$ and $\delta _{AB}$ can be
determined with great accuracy: $\gamma _{1}$=404$\pm $10meV and $\delta
_{AB}$=18$\pm $2meV. Therefore, the $\sigma _{1}\left( \omega ,V\right) $
data at low biases clearly indicates an asymmetry between valence and
conduction bands in bilayer graphene due to finite energy difference of A
and B sublattices.

In order to explore the V dependence of $\omega _{peak}$ and the width of
the main peak in $\sigma _{1}\left( \omega ,V\right) ,$ $\Gamma _{peak}$, we
plot the e$_{2}$ and e$_{3}$ transition energies \cite{Nicol} as a function
of V (Fig. 3b), using the gap formula $\Delta $(V) in \cite{McCann}\cite%
{note1} and our calculated asymmetric dispersion E$_{i}$(k) (i=1,2,3,4) \cite%
{note2}, with v$_{F}$=1.1$\times $10$^{6}$m/s, $\gamma _{1}$=404meV, $\delta
_{AB}$=18meV, and for both v$_{4}=0$ and v$_{4}=0.04$. We find that e$_{2}$
does not depend on v$_{4}$ \cite{note1}, whereas e$_{3}$ is strongly
affected by v$_{4}$. With a finite value of v$_{4}$ ($\approx $0.04), an
assignment of $\omega _{peak}$ to (e$_{2}$+e$_{3}$)/2 appears to fit our
data well on both electron and hole sides. Nevertheless, larger separation
of e$_{2}$ and e$_{3}$ on the hole side is inconsistent with the relatively
narrow peak in $\sigma_{1}\left( \omega ,V\right) $ for both electron and
hole injection with nearly identical width. Yet the finite value of the v$%
_{4} $ parameter is essential to qualitatively account for the voltage
dependence of $\omega _{peak},$ because with v$_{4}$$\approx$0 $\omega
_{peak}$\ follows e$_{2}$ and e$_{3}$ on the electron and hole sides (Fig
3b), respectively, eluding a consistent description. A quantitative
understanding of the V dependence of $\omega _{peak}$ and $\Gamma _{peak}$
is lacking at this stage. Our results highlight a need for further
experimental and theoretical investigation of v$_{4}$ including its possible
V dependence.

We stress that $\gamma _{1}$ and $\delta _{AB}$ are determined from the low
bias (low k$_F$) data. Therefore the values of $\gamma _{1}$ and $\delta
_{AB}$ reported here do not suffer from the currently incomplete
understanding of V dependence of $\omega _{peak}$ and $\Gamma_{peak}$
discussed above. The $\gamma _{1}$ value (404$\pm $10meV) is directly
determined from measurements of transitions between the two conduction bands
or valence bands. It has been predicted theoretically that the band
structure of bilayer graphene as well as the parameters $\gamma _{1}$ and v$%
_{F}$ can be strongly renormalized by electron-electron interactions \cite%
{Kusminskiy}. The $\gamma _{1}$ value inferred from our data is close to
theoretical estimates of the renormalized $\gamma _{1}$ \cite{Kusminskiy}.

IR measurements reported here have enabled accurate extraction of $\delta
_{AB}$ in bilayer graphene free from ambiguities of alternative experimental
methods. Interestingly, the energy difference between A and B sublattices $%
\delta _{AB}$ in bilayer graphene (18meV) is much greater than that in
graphite ($\delta _{AB}\approx $8meV) \cite{Graphite}. Such a large value of 
$\delta _{AB}$ in bilayer is very surprising. We propose that this
observation stems from different interlayer coupling between the B
sublattices in bilayer graphene and graphite. In bilayer, the direct
interlayer coupling between A$_{1}$ and A$_{2}$ (Fig 1(b)) considerably enhances the
energy of A sublattices due to Coulomb repulsion between the $%
\pi $ orbits. However, the sublattices B$_{1}$ and B$_{2}$ are not on top of
each other as shown in Fig 1(b) and thus are more weakly coupled. Therefore, the
energy of the B sites is lower than that of the A sites within the same
layer, leading to a large $\delta _{AB}$ in bilayer graphene. On the other
hand, in graphite the B sublattice in the third layer B$_{3}$ is on top of B$%
_{1}$, and that in the fourth layer B$_{4}$ is right above B$_{2}$. The
coupling between the B sites in the next nearest neighbor layers (B$_{1}$
and B$_{3}$, B$_{2}$ and B$_{4}$, etc) increases the energy of the B sites
compared to that in bilayer, giving rise to a smaller $\delta _{AB}$ value.

The asymmetry between valence and conduction bands uncovered by our study
has broad implications on the fundamental understanding of bilayer graphene.
An electron-hole asymmetry was observed in the cyclotron resonance \cite%
{Erik} and cyclotron mass experiments \cite{Castro} in bilayer, both of
which have eluded a complete understanding so far. Our accurate
determination of finite values of $\delta _{AB}$ and v$_{4}$ calls for
explicit account of the asymmetric band structure in the interpretation of
the cyclotron data. Moreover, the different $\delta _{AB}$ values in bilayer
graphene and graphite reveal the importance of interlayer coupling in
defining the electronic properties and band structure of graphitic systems.

During the preparation of this paper, we became aware of another infrared
study of bilayer graphene by A.B. Kuzmenko et al \cite{Kuzmenko-bilayer}. We
thank M.L. Zhang and M.M. Fogler for fruitful discussions. Work at UCSD is
supported by DOE (No. DE-FG02-00ER45799). Research at Columbia University is
supported by the DOE (No. DE-AIO2-04ER46133 and No. DE-FG02-05ER46215), NSF
(No. DMR-03-52738 and No. CHE-0117752), NYSTAR, the Keck Foundation and
Microsoft, Project Q. The Advanced Light Source is supported by the
Director, Office of Science, Office of Basic Energy Sciences, of the U.S.
Department of Energy under Contract No. DE-AC02-05CH11231.

\begin{figure}[tbp]
\includegraphics[width=6cm, height=8.3cm]{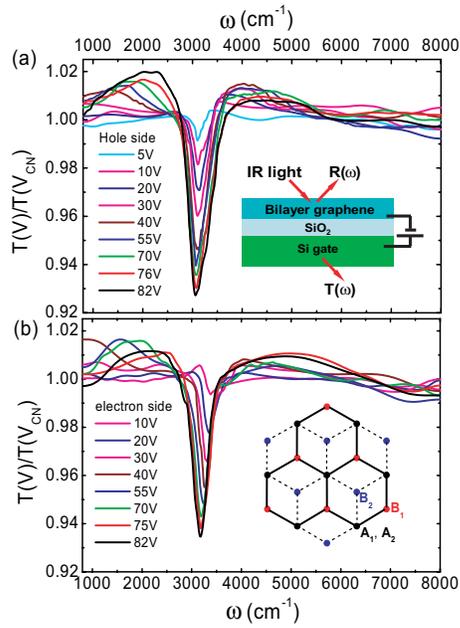}
\caption{(color online) T(V)/T(V$_{CN}$) spectra of bilayer graphene. (a)
and (b): data for E$_{F}$ on the hole side and electron side. Inset of (a):
a schematic of the device and infrared measurements. Inset of (b): a
schematic of bilayer graphene. The solid lines indicate bonds in the top
layer A$_{1}$B$_{1}$, whereas the dashed lines indicate bonds in the bottom
layer A$_{2}$B$_{2}$. The sublattice A$_{1}$ is right on top of the
sublattice A$_{2}$.}
\label{Fig.1}
\end{figure}

\begin{figure}[tbp]
\includegraphics[width=6cm, height=12.5cm]{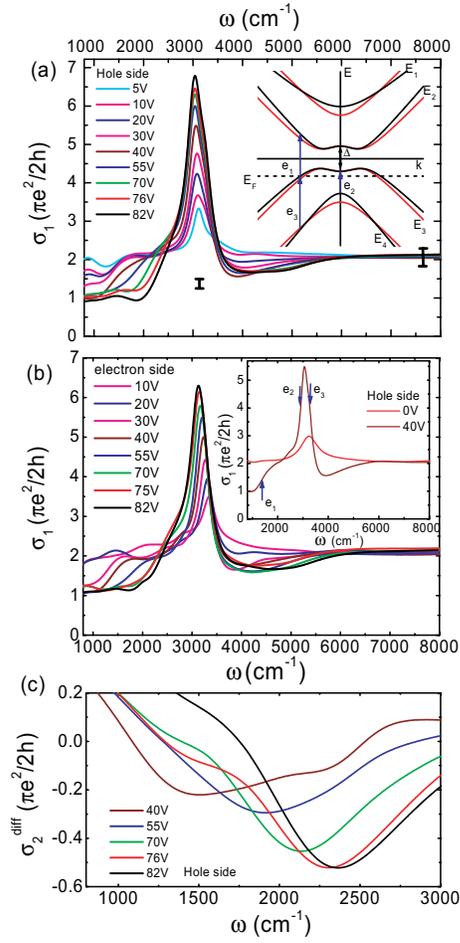}
\caption{(color online) The optical conductivity of bilayer graphene. (a)
and (b): $\protect\sigma _{1}\left( \protect\omega ,V\right) $ data for E$%
_{F}$ on the hole side and electron side. (c): $\protect\sigma %
_{2}^{diff}\left( \protect\omega ,V\right) $ spectra in the low frequency
range, after subtracting the Lorentzian oscillators describing the main
resonacne around 3000 cm$^{-1}$ from the whole $\protect\sigma _{2}\left( 
\protect\omega ,V\right) $ spectra. Inset of (a): Schematics of the band
structure of bilayer with zero values of $\protect\delta _{AB}$ and v$_{4}$
(red) and finite values of $\protect\delta _{AB}$ and v$_{4}$ (black),
together with allowed interband transitions. Insets of (b): $\protect\sigma %
_{1}\left( \protect\omega ,V\right) $ at 0V (V$_{CN}$) and 40V on the hole
side with assignments of the features. }
\label{Fig.2}
\end{figure}

\begin{figure}[tbp]
\includegraphics[width=6cm, height=7.18cm]{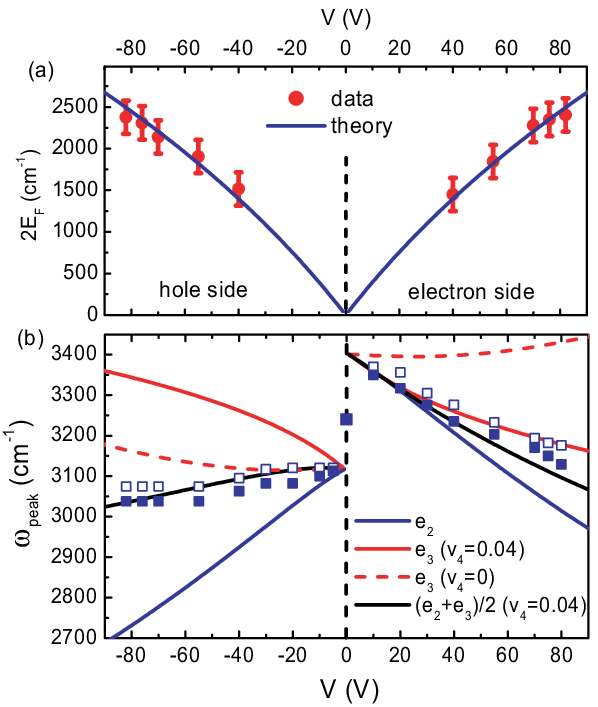}
\caption{(color online) (a) Symbols: the 2E$_{F}$ values extracted from the
optical conductivity detailed in the text. The error bars are estimates of
the uncertainties of $\protect\sigma _{2}^{diff}\left( \protect\omega %
,V\right) $ spectra in Fig 2(c). Solid lines: the theoretical 2E$_{F}$
values using v$_{F}$=1.1$\times $10$^{6}$ m/s and $\protect\gamma _{1}$%
=450meV. (b) Solid symbols, the energy of the main peak $\protect\omega %
_{peak}$ in the $\protect\sigma _{1}\left( \protect\omega ,V\right) $
spectrum. Open symbols: the energy of the dip feature $\protect\omega _{dip}$
in the T(V)/T(V$_{CN}$) spectra. Note that $\protect\omega _{peak}$ in $%
\protect\sigma _{1}\left( \protect\omega ,V\right) $ is shifted from $%
\protect\omega _{dip}$ in the raw T(V)/T(V$_{CN}$) data with an almost
constant offset, which is due to the presence of the substrate. Solid lines:
theoretical values of the transitions at e$_{2}$, e$_{3}$ and (e$_{2}$+e$_{3}
$)/2 with v$_{F}$=1.1$\times $10$^{6}$m/s, $\protect\gamma _{1}$=404meV and $%
\protect\delta _{AB}$=18meV and v$_{4}$=0.04. Red dashed lines: e$_{3}$ with
similar parameters except v$_{4}$=0.}
\label{Fig.3}
\end{figure}


\begin{thebibliography}{99}
\bibitem{note1} The gap formula $\Delta $(V) in \cite{McCann} did not take
into account $\delta _{AB}$ and v$_{4}$. However, we find that finite\ $%
\delta _{AB}$ and v$_{4}$ values have no effect on the gap $\Delta $.
Specifically, $\delta _{AB}$ primarily modifies the E$_{1}$ and E$_{4}$
bands, while leaving unchanged  the gap $\Delta $ between the E$_{2}$ and E$%
_{3}$ bands at k=0, as shown in Fig 2(a). In addition, v$_{4}$ always
appears in a term v$_{4}$k in the Hamiltonian\cite{Nilsson}, therefore it
has zero effect on $\Delta $.
\bibitem{note2} In the calculation of the energy bands E$_{i}$(k)
(i=1,2,3,4), we used the approximation $\Delta $=0, which can be justified
for the purpose of estimating e$_{2}$ and e$_{3}$. The gap $\Delta $ is very
small ($<$80 meV) in the voltage range studied in our work \cite{McCann},
and does not affect the higher energy bands E$_{1}$(k=0) or E$_{4}$(k=0) and
therefore the value of e$_{2}$. Note that $\Delta $=0 is only assumed when
calculating E$_{i}$(k), but not in the $\Delta $ term in the expression of e$%
_{2}$. Moreover, E$_{2}$ and E$_{3}$ bands are modified by the gap only at
energies below $\Delta $/2. Because E$_{F}$ is much larger than $\Delta $/2
under applied voltage \cite{McCann}, E$_{2}$(k=k$_{F}$)and E$_{3}$(k=k$_{F}$%
) are not affected by $\Delta$. Therefore, a finite gap does not modify the
value of e$_{3}$ compared to that with $\Delta $=0.
\end{thebibliography}
\end{document}